\documentclass[preprint,showpacs,preprintnumbers,amsmath,amssymb]{revtex4}
\usepackage{graphicx}
\usepackage{dcolumn}
\usepackage{bm}
\usepackage{natbib}
\usepackage{amsthm}

\def\vx{\mathbf{x}}
\def\vF{\mathbf{F}}
\def\vJ{\mathbf{J}}

\begin{document}

\preprint{}

\title{Stochastic Thermodynamics Across Scales: Emergent 
Inter-attractoral Discrete Markov Jump Process and Its 
Underlying Continuous Diffusion}

\author{Mois\'es Santill\'{a}n}
%\email{msantillan@cinvestav.mx}
\affiliation{Centro de Investigaci\'{o}n y Estudios Avanzados del IPN, Unidad Monterrey, Parque de Investigaci\'{o}n e Innovaci\'{o}n Tecnol\'{o}gica, 66600 Apodaca NL, M\'{E}XICO}

\author{Hong Qian}
%\email{qian@amath.washington.edu}
\affiliation{Department of Applied Mathematics, University of Washington, Box 352420, Seattle, WA 98195,
USA}

\begin{abstract}
The consistency across scales of a recently developed mathematical thermodynamic structure, between a continuous stochastic nonlinear dynamical system (diffusion process with Langevin or Fokker-Planck equations) and its emergent discrete, inter-attractoral Markov jump process, is investigated.  We analyze how the system's thermodynamic state functions, e.g. free energy $F$, entropy $S$, entropy production $e_p$, and free energy dissipation $\dot{F}$, etc., are related when the continuous system is describe with a coarse-grained discrete variable. We show that the \emph{thermodynamics} derived from the underlying detailed continuous dynamics is \emph{exact} in the Helmholtz free-energy representation. That is, the system thermodynamic structure is the same as if one only takes a middle-road and starts with the ``natural'' discrete description, with the corresponding transition rates empirically determined. By ``natural'', we mean in the thermodynamic limit of large systems in which there is an inherent separation of time scales between inter- and intra-attractoral dynamics. This result generalizes a fundamental idea from chemistry and the theory of Kramers by including thermodynamics: while a mechanical description of a molecule is in terms of continuous bond lengths and angles, chemical reactions are phenomenologically described by the Law of Mass Action with rate constants, and a stochastic thermodynamics.   
\end{abstract}


\maketitle

\section{Introduction}

Recently, a quite complete mathematical thermodynamic structure for general stochastic processes has been proposed, for both discrete Markov jump processes and continuous Langevin-Fokker-Planck systems \cite{Ge:2010fk,esposito_prl:2010,qian_decomp}. The entropy production rate $e_p$ of a Markov dynamics can be mathematically decomposed into two non-negative terms: free energy dissipation rate $-\dot{F}$, corresponding to Boltzmann's original theory on irreversibility of spontaneous change, and house-keeping heat $Q_{hk}$, corresponding to Brussels school's notion of irreversibility in nonequilibrium steady states (NESS) \cite{NP_book,zqq,gqq}. For almost all applications of stochastic dynamic theories in physics, chemistry and biology, there will be multiple time scales, and often with a significant separation. When a dynamical system is highly nonlinear, and its interaction network includes feedbacks, multistability with several attractors is often the rule rather than exception.  On the other hand, the concept of ``landscape'' has become a highly popular metaphor as well as a useful analytical device \cite{wolynes,ge_qian_lc}. When stochastic nonlinear dynamical systems of populations of individuals become large, a time scale separation between inter- and intra-attractoral dynamics becomes almost guaranteed. In cellular biology, they have been called {\em biochemical network} and {\em cellular evolution} time scales
respectively \cite{qian_iop}.

In chemistry, a separation of time scales has lead to a fundamental understanding of chemical reactions in terms of discrete states of molecules, in complementary to the full mechanical description of constitutive atoms in terms of bond lengths and bond angles. In fact, one of the most significant, novel chemical concepts is ``transition state'', which in terms of modern nonlinear dynamical systems is the saddle point on a separatrix that divides two attractors \cite{wolynes_sfi}. Recall also that in applications of Gibbs' formalism of statistical mechanics to chemical equilibrium, the conditional free energy plays a central role \cite{ben_naim_book,bwt_book}. One usually does not work with the pure mechanical energy of a system; rather, one works with a {\em conditional free energy} from a coarse-grained representation and develops a partition function thereafter. An essential notion in this approach is the {\em consistency across scales}. We shall expand on these ideas more precisely in the following section. 

In the present work we address the question of ``whether the mathematical thermodynamic structure of a given continuous stochastic nonlinear dynamical system is consistent with the one associated with the emergent discrete Markov jump process.'' In other words: whether the formal mathematical relations between state functions and process variables remain unchanged when the system is viewed at either a finer- or a coarse-grained scale. 

It is important to point out, at the onset, that the ``state'' of a stochastic dynamical system has always had two distinctly different meanings: ($a$) a state of a single, stochastically fluctuating, system; and ($b$) a state in terms of the distribution over an ensemble.  In more precise mathematical terms, ($a$) are functions of a stochastic process, while ($b$) are functionals of the solution to a Fokker-Planck equation.  The deep insight from the theory of 
probability is that these are two complementary, yet mathematically
identical, descriptions of a same dynamical process.  With this distinction in mind, entropy and free energies are state functionals of the second type, while energy is a state function of the first type. A state function of the first type naturally has fluctuations. On the other hand, most classical thermodynamic functions are the second type. 

Attempting to introduce entropy as a function of the first type, \citet{Qian:2001fk} defined a trajectory based entropy $\Upsilon_t = -\ln f_X^s(X_t)$ where $X_t$ is a diffusion process, and $f_X^s(x)$ is the stationary solution to the corresponding Fokker-Planck (Kolmogorov forward) equation.  One immediately sees that entropy is really a population-based concept. For irreversible diffusion processes (i.e., without detailed balance),  $f_X^s(x)$ is non-local \cite{Qian:2001fk}. However, for reversible diffusion processes with detailed balance, since $f^s_X(x)\propto e^{-\phi(x)}$ where $\phi(x)$ is potential energy, fluctuating $\Upsilon_t$ and fluctuating energy $\phi(X_t)$ are the same.

\section{Equilibrium statistical thermodynamic consistency across scales}

In equilibrium statistical mechanics, the concept of \emph{consistency}---or invariance---has a fundamental importance in the study of realistic physical systems at an appropriate scale \cite{bwt_book,ben_naim_book}. In a continuous system, the conditional free energy is known as the {\em potential of mean force} \cite{Kirkwood:1935fk}. The conditional free energy can do work just as the Newtonian mechanical energy; the concept of {\em entropic force} is well understood in physical chemistry \cite{dill_book}.

For an investigator working on certain level of description, with discrete states ($i=1,2,\cdots$) and conditional free energy ($A_i$), the canonical partition function of the statistical thermodynamic system is \cite{bwt_book,ben_naim_book,dill_book}
\begin{equation}
Z(T) = \sum_{i=1} e^{-A_i/k_BT}.
\label{eq_01}
\end{equation}
Note that, since $A_i$ is a conditional free energy, it can be decomposed into $A_i = E_i-TS_i$, where $E_i =\partial (A_i/T)/\partial (1/T)$ and $S_i=-\partial A_i/\partial T$.  In general, both $E_i$ and $S_i$ are themselves functions of the temperature.

Now, for another investigator who works at a much more refined level, with a continuos variable $\vx$, each state $i$ corresponds to a unique region of the phase space $\omega_i$, with $\omega_i\bigcap\omega_j=\emptyset$ for $i\neq j$, and $\bigcup_{i=1} \omega_i = \Omega$ covering the entire phase-space region available to the system.  Let $V(\vx)$ ($\vx\in\Omega$) be the potential of mean force at this level. Then, his/her canonical partition function is
\begin{equation}
	\widetilde{Z}(T) = \int_{\Omega} d\vx e^{-V(\vx)/k_BT}.
\label{eq_02}
\end{equation}

We see that $Z(T)$ and $\widetilde{Z}(T)$ are equal if the $A_i$ in Eq. (\ref{eq_01}) are such that
\begin{equation}
	A_i(T)=-k_BT\ln\left(\int_{\omega_i} d\vx e^{-V(\vx)/k_BT}
				\right).
\label{eq_03}
\end{equation}
The equality $Z(T)=\widetilde{Z}(T)$ following Eqs. (\ref{eq_01}) and (\ref{eq_03}) is exact in equilibrium statistical mechanics. Nonetheless, as we demonstrate in this work, its generalization to include dynamics requires a separation of time scales for the dynamics within each $\omega_i$ and the dynamics between $\omega$'s (this is well understood in physical chemistry as ``rapid equilibrium'' averaging).  Here, we choose the $\omega$'s according to the basins of attraction of the underlying nonlinear dynamics. In this case, the separation of time scales for intra- and inter-attractoral dynamics is widely accepted. 

The mathematical origin of the consistency discussed above relies in fact upon the concepts of \emph{conditional probability}, \emph{marginal probability}, and the \emph{law of total probability}!  The free energy  of a system, or a sub-system, is directly related to its probability. The same cannot be said for the entropy \cite{noyes61,Qian:1996uq}, which increses with more detailed descriptions and is also coordinate-system dependent for continuous variabels:
\begin{eqnarray}
	S(T) &=& -k_B\sum_i\left(\frac{e^{-A_i/k_BT}}{Z(T)}\right)
		\ln\left(\frac{e^{-A_i/k_BT}}{Z(T)}\right)
\nonumber\\
	&\leq& -k_B\int_{\Omega} d\vx \left(\frac{e^{-V(\vx)/k_BT}}
		{\widetilde{Z}(T)}\right)
		\ln\left(\frac{e^{-V(\vx)/k_BT}}{\widetilde{Z}(T)}\right)
		  = \widetilde{S}(T). 
\label{EntRef}
\end{eqnarray}
The proof for the inequality can be found in any text on information theory \cite{cover_book}.  Also see Appendix \ref{appendix}.

Note that since internal energy is the sum $-k_BT\ln Z(T) + TS(T)=E(T)$, one immediately has $\widetilde{E}(T)\ge E(T)$ across scales as well. This leads to a type of entropy-energy compensation \cite{Qian:1996uq, Qian:1998kx, Qian:2001fk, Santillan:2011vn}.

\section{Open System concepts and definitions}

\subsection{Fokker-Planck equation, stationary distribution, and detailed balance}

Consider a system whose state is represented by variable $\vx$, and assume that $\vx$ is a stochastic variable following a continuous-space continuous-time diffusion process. Let $P(\vx,t)$ denote the probability density of finding the system in state $\vx$ at time $t$. In what follows we shall assume that the master equation (Chapman-Kolomogorov equation) governing the dynamics of $P(\vx,t)$ can be represented by the following Fokker-Planck equation \footnote{Following \citeauthor{Kubo:1973fk} we write the Fokker-Planck equation for the probability density function of a continuous diffusion in the divergence form. Other choices such as Ito or Stratonovich forms can be readily transformed into the present one.}:
\begin{equation}
\frac{\partial P(\vx,t)}{\partial t} = -\nabla \cdot \vJ,
\label{fpe}
\end{equation}
where 
\begin{equation}
\vJ(\vx,t) = -D(\vx)\left[\epsilon\nabla P(\vx,t) 
           + \vF(\vx) P(\vx,t)\right]
\label{probcurr}
\end{equation}
is the probability current. In equation (\ref{probcurr}), $D(\vx)$ is the diffusion coefficient, $\vF(\vx)$ is the force (not necessarily conservative) acting upon the system, and $\epsilon$ is a parameter which will serve as our ``temperature''.  For fluctuations of isothermal molecular systems in equilibrium at temperature $T$, Einstein's relation dictates that $\epsilon= k_BT$, where $k_B$ is Boltzmann's constant.  However, in the present work, the notion of temperature does not exist.

We shall assume that the system can be driven and approaches to a nonequilibrium steady state in infinite time \cite{zqq,gqq}.  The nonequilibrium driving force comes from a ``chemical driving force'' in $\nabla\times\vF(\vx)\neq 0$ \cite{qian_jpc_06, qian_arpc}.  When $\vF(\vx)$ is conservative,
\[
        \nabla\times \vF(\vx)= 0 \ 
        \Rightarrow \
        \vF(\vx)=-\nabla V(\vx).
\]
Then, the stationary $P^{s}(\vx)=e^{-V(\vx)/\epsilon}$, while the stationary $\vJ(\vx)=0$. Furthermore, the stationary distribution $P^{s}(\vx)$ complies with detailed balance. That is, $P^{s}(\vx)$ is analogous to thermodynamic equilibrium \citep{Qian:2002fk}. Hence, a stationary system (\ref{fpe}) is also mathematically called equilibrium in this case \cite{zqq,gqq}.

Let us assume that Eq. (\ref{fpe}) has one single ergodic stationary solution $P^s(\vx)$ with the corresponding stationary current $\nabla\cdot\vJ^s(\vx)=0$; but usually, $\vJ^s(\vx)\neq 0$. We shall again write the stationary probability density as
\begin{equation}
P^s(\vx) = C \exp \left(-\Psi_{\epsilon}(\vx)/\epsilon \right),
\label{IrrevPot}
\end{equation}
where function $\Psi_{\epsilon}(\vx)$ is known as the \emph{non-equilibrium} potential \cite{Kubo:1973fk}, $C = \left[ \int_{\Omega} d\vx \exp \left(- \Psi_{\epsilon}(\vx) /\epsilon\right) \right]^{-1}$ is a normalization constant, and $\Omega$ represents the region of the state space available to the system.  Note that in general $\Psi_{\epsilon}(\vx)$ is actually also a function of $\epsilon$.  However, for many interesting applications, $\Psi_{\epsilon}(\vx)$ is a function of $\vx$ alone in the limit of $\epsilon\rightarrow 0$. The probability current $\vJ^s(\vx)$ is a time-invariant, divergence-free vector field for the 
stationary distribution $P^s(\vx)$ \cite{qian_decomp}.

\subsection{Small $\epsilon$ limit}

Let us consider first the case where $\vF(\vx)$ is conservative. When $\epsilon=0$, the system dynamic behavior is dictated, in a deterministic fashion, by the potential $V(\vx)$. That is, depending on the initial condition, the system state will evolve towards one of the local minima of $V(\vx)$ and will remain there indefinitely. In that sense, every local minimum of $V(\vx)$ corresponds to a stable steady state. Moreover, each stable steady state has a basin of attraction associated to it.  Whenever the initial condition lies within a given basin of attraction, the system will eventually reach the state corresponding to the local minimum $V(\vx)$ point. Finally, all neighboring basins of attraction are separated by saddle points and separatrices which the system has to surpass in order to go from one basin to the other.

If $\epsilon$ is not zero, but very small as compared to the height of the saddle points and separatrices between basins of attraction, the stationary probability distribution $P^s(\vx)$ will present high narrow peaks around the stationary states, and will attain very low values at the saddle nodes separating neighboring attractive basins. This further implies that the transition rates between every two attractive basins are small as well, as compared with the probability relaxation-rates within each basin.

In the case of a non-conservative force, the stationary distribution depends on the non-equilibrium potential $\Psi_{\epsilon}(\vx)$, when it exists, in an analogous  way as $P^s(\vx)$ depends on $V(\vx)$ for conservative forces \cite{fw_book}. This means that the above considerations could be still valid when $\vF(\vx)$ is non conservative. In particular, $\Psi_{\epsilon}(\vx)$ defines a landscape in the state space \cite{ge_qian_lc}, a basin of attraction can be identified around each of the local minima of $\Psi_{\epsilon}(\vx)$ and, in the small $\epsilon$ limit, $P^s(\vx)$ presents high narrow peaks around each minimum of $\Psi_{\epsilon}(\vx)$ and takes very low values at the saddle points and separatrices that separate neighboring attractive basins.  See \cite{ge_qian_lc} for systems with limit cycles.

\section{Probability discretization}

\subsection{Discretization of the state space}

Consider a system whose non-equilibrium potential $\Psi_{\epsilon}(\vx)$ has $N$ local minima with the corresponding basins of attraction in the state space. Let $\omega_i$ be the region of the state space delimited by the attractive basin of the $i$th local minimum of $\Psi_{\epsilon}(\vx)$, and let $\Xi_i$ denote the boundary of $\omega_i$. In Appendix \ref{boundary} we demonstrate that $\Xi_i$ can always be written as
\begin{equation}
\Xi_i = \bigcup_{j=0}^N \Xi_{ij},
\label{boundaries}
\end{equation}
where $\Xi_{ij}=\Xi_{ji}$ ($j=1,2\dots N$) represents the common boundary between $\omega_i$ and $\omega_j$, while $\Xi_{i0}$ denotes the part of the $\omega_i$ boundary not shared with any other region. In case that $\omega_i$ and $\omega_j$ share no boundary, $\Xi_{ij} = \emptyset$.

From the above considerations, the probability $P_i$ that the system state is in region $\omega_i$ is
\begin{equation}
P_i(t) = \int_{\omega_i} d\vx P(\vx,t).
\label{pi}
\end{equation}
Furthermore, it follows from (\ref{fpe}) and Stokes' theorem that
\begin{equation}
\frac{d P_i(t)}{dt} = \int_{\omega_i} d\vx \frac{\partial P(\vx,t)}{\partial t} 
= - \sum_{j=1}^N \int_{\Xi_{ij}} d\mathbf{s} \cdot \vJ(\vx,t) .
\label{dpidtaux}
\end{equation}
In the derivation of the above equation we have assumed that the probability current $\vJ$ is zero along $\Xi_{i0}$.

Let us analyze the integral $\int_{\Xi_{ij}} d\mathbf{s} \cdot \vJ$. From (\ref{probcurr}), it can be rewritten as
\begin{eqnarray}
\int_{\Xi_{ij}} d\mathbf{s} \cdot \vJ & = & - \int_{\Xi_{ij}} d\mathbf{s} \cdot [\epsilon D(\vx) \nabla P(\vx,t)] \; \mathrm{H}(- d\mathbf{s} \cdot [\epsilon D(\vx) \nabla P(\vx,t)])  \nonumber \\
 &  & -  \int_{\Xi_{ij}} d\mathbf{s} \cdot [\epsilon D(\vx) \nabla P(\vx,t)] \nonumber \; \mathrm{H}(d\mathbf{s} \cdot [\epsilon D(\vx) \nabla P(\vx,t)]) \\
 &  & + \int_{\Xi_{ij}} d\mathbf{s} \cdot [D(\vx) \mathrm{F}(\vx) P(\vx,t)] \; \mathrm{H}( d\mathbf{s} \cdot [D(\vx) \mathrm{F}(\vx) P(\vx,t)]) \nonumber \\
 &  & +  \int_{\Xi_{ij}}  d\mathbf{s} \cdot [D(\vx) \mathrm{F}(\vx) P(\vx,t)] \; \mathrm{H}( - d\mathbf{s} \cdot [D(\vx) \mathrm{F}(\vx) P(\vx,t)]) , \nonumber
\end{eqnarray}
with $\mathrm{H}(\cdot)$ being Heaviside's function. Given that $\mathrm{H}(x)>0$ if and only if $x>0$, just one of the first two terms in the right hand side of the previous equation is positive or zero, while the other is negative or zero; the same is true for the last two terms. Let us define $J_{ij}$ as the sum of the two positive terms, and $J_{ji}$ as minus the sum of the two negative terms. Hence,
\begin{equation}
\int_{\Xi_{ij}} d\mathbf{s} \cdot \vJ =  J_{ij} - J_{ji}.
\label{sumcurr}
\end{equation}
From its definition, $J_{kl} \geq 0$ for all $k,l = 1,2\dots N$. Furthermore, from Eq. (\ref{sumcurr}), $J_{kl}$ can be interpreted as the net probability flux from $\omega_k$ into $\omega_l$. Finally, by substituting Eq. (\ref{sumcurr}) into Eq. (\ref{dpidtaux}) we obtain
\begin{equation}
\frac{d P_i(t)}{dt} =  \sum_{j=1}^N \left(J_{ji} - J_{ij}\right).
\label{dpidt}
\end{equation}

\subsection{Adiabatic approximation and Kramers theory}

Following \citet{Kampen:2007kx}, \citet{Risken:1996uq} and
Freidlin-Wentzell \cite{fw_book} we make use of the assumption that $\epsilon$ is much smaller than the height of the saddle nodes between every two attractive basins so that the corresponding transition rates are very small, as compared with the probability dynamics inside each basin. In consequence, the probability distribution within any $\omega_i$ can be approximated by the quasi-stationaty distribution 
\begin{equation}
P(\vx,t) \approx C_i(t)\exp(-\Psi_{\epsilon}(\vx)/\epsilon),
\label{qsdist}
\end{equation}
with $C_i(t)$ given by
\begin{equation}
C_i(t) = \frac{ P_i(t) }{\int_{\omega_i} d \vx \exp(-\Psi_{\epsilon}(\vx)/\epsilon)  },
\label{normconst}
\end{equation}
so that $\int_{\omega_i} d \vx P(\vx,t)  = P_i(t) $. From the approximation above, and a theorem from \cite{fw_book} that justifies the application of Kramers' theory to any pair of
adjacent $i$ and $j$ \citep{Kampen:2007kx,Risken:1996uq}, it follows that
\begin{equation}
J_{ij}(t) = \gamma_{ij} P_i(t),
\label{linearflux}
\end{equation}
where the transition rates $\gamma_{ij}$ are determined by the 
so-called local pseudo-potential; particularly, by the height 
of the saddle points between neighbouring $i$ and $j$ attractors \cite{weinan,ge_qian_lc}.

Finally, by substituting Eq. (\ref{linearflux}) into Eq. (\ref{dpidt}) we obtain the following master equation for $P_i(t)$:
\begin{equation}
\frac{d P_i(t)}{dt} =  \sum_{j=1}^N \gamma_{ji} P_j(t) - \gamma_{ij} P_i(t).
\label{me}
\end{equation}
Note that, since in the stationary state $J_{ij}^s=P^s_i\gamma_{ij}$, but $J^s_{ij}\neq J^s_{ji}$ in general, $P^s_i\gamma_{ij} \neq P^s_j\gamma_{ji}$.  Therefore the emergent master equation in (\ref{me}) is not necesarilly detail balanced.

\section{Thermodynamic state functionals}

\subsection{Internal Energy}

Under the assumptions that the system modeled by Eq. (\ref{fpe}) has a unique stationary distribution one can mathematically define, following Kubo \cite{Kubo:1973fk} and many others including Ge and Qian \cite{Ge:2010fk}, the energy function associated to state $\vx$ via the stationary distribution $P^s(\vx)$ as
\begin{equation}
\phi(\vx) = -\epsilon \ln P^s(\vx).
\label{udef}
\end{equation}
In systems with detailed balance, $P^s(\vx)$ equals the thermodynamic-equilibrium probability distribution $P^e(\vx)$ and Eq. (\ref{udef}) is equivalent to Boltzmann's law---provided we choose the zero level of free energy such that the partition function equals one. When detailed balance is not fulfilled, Kubo et. al. called  $\phi(\vx)$ a \emph{stochastic potential} \citep{Kubo:1973fk}.  Finally, from (\ref{udef}), the mean ``energy'' of the mesoscopic state $P(\vx,t)$ can be written as
\begin{equation}
U(t) = \int_{\Omega} d \vx P(\vx,t)\phi(\vx) = -\epsilon \int_{\Omega} d \vx P(\vx,t) \ln P^s(\vx).
\label{IntEner}
\end{equation}

Given the definition of the attractive basins, $\Omega = \bigcup_{i=1}^N \omega_i$, while $\omega_i \bigcap \omega_j = \emptyset$ for all $i \neq j$. Then, Eq. (\ref{IntEner}) can be rewritten as 
\begin{equation}
U(t) = -\epsilon\sum_{i=1}^N P_i(t) \ln P_i^s -\epsilon \sum_{i=1}^N P_i(t) \int_{\omega_i} d \vx\frac{P(\vx,t)}{P_i(t)} \ln \frac{P^s(\vx)}{P_i^s},
\label{udecomp}
\end{equation}
with $ P_i^s = \int_{\omega_i} d \vx P^s(\vx)$. Finally, substitution of Eqs. (\ref{qsdist}) and (\ref{normconst}) into Eq. (\ref{udecomp}) leads to
\begin{equation}
U(t) = -\epsilon\sum_{i=1}^N P_i(t) \ln P_i^s +\sum_{i=1}^N P_i(t) \tilde{s}_i,
\label{udecomp2}
\end{equation}
where
\begin{equation}
\tilde{s}_i = -\epsilon \int_{\omega_i} d \vx \frac{\exp(-\phi(\vx/\epsilon))}{Z_i} \ln \frac{\exp(-\phi(\vx/\epsilon))}{Z_i},
\label{tildes}
\end{equation}
and ${Z_i} = \int_{\omega_i} d \vx \exp(-\phi(\vx/\epsilon))$ . The first term in the right hand side of Eq. (\ref{udecomp2}) can be interpreted as a coarse-grained contribution to the system's internal energy, arising from the distribution of probability among the $N$ available attractive basins. On the other hand, the second term in the right hand side of Eq. (\ref{udecomp2}) corresponds to the fine-grained contribution to the system internal energy, due to the distribution of probability density $P(\vx,t)$ within each basin. The Boltzmann-like form of the terms within the integral originates from the adiabatic approximation we have made.

\subsection{Entropy and Free Energy}

The Gibbs entropy is defined as usual:
\begin{equation}
S(t) = -\epsilon \int_{\Omega} d \vx P(\vx,t) \ln P(\vx,t) .
\label{Entropy}
\end{equation}
By following an analogous procedure to the one in the previous subsection, known as the chain rules for entropy and 
relative entropy, Eq. (\ref{Entropy}) can be rewritten as
\begin{eqnarray}
S(t) & = & -\epsilon\sum_{i=1}^N P_i(t) \ln P_i^s -\epsilon\sum_{i=1}^N P_i(t) \int_{\omega_i} d \vx \frac{P(\vx,t)}{P_i(t)} \ln \frac{P(\vx,t)}{P_i(t)} , \nonumber \\
 & = &-\epsilon\sum_{i=1}^N P_i(t) \ln P_i(t) + \sum_{i=1}^N P_i(t) \tilde{s}_i,
\label{sdecomp}
\end{eqnarray}
Once again, the entropy can be decomposed into a coarse-grained contribution---due to the distribution of probability among the $N$ available attractive basins---as well as a fine-grained contribution due to the distribution of the probability density $P(\vx,t)$ within each basin. This result is in complete agreement with that in Eq. (\ref{EntRef}). Notice that, because of the adiabatic approximation, the fine-grained contributions to both $U$ and $S$ happen to be equal.

From its definition, the mean Helmholtz free energy is
\begin{equation}
F(t) = U(t)-S(t) = \epsilon \int_{\Omega} d \vx P(\vx,t) \ln \frac{P(\vx,t)}{P^s(\vx)} .
\label{FreeEnergy}
\end{equation}

Furthermore, after performing the separation into coarse- and fine-grained contributions we obtain
\begin{equation}
F(t) = \epsilon \sum_{i=1}^N P_i(t) \ln \frac{P_i(t)}{P_i^s}.
\label{fdecomp}
\end{equation}
Observe that, in this case, the fine-grained contribution is absent, the reason being that the corresponding terms in $U$ and $S$ cancel at the time of subtracting.

\subsection{Further thermodynamic significance underlying the scale separation}

In the coarse-grained perspective one can define  
\begin{equation}
u_i = - \epsilon \ln P_i^s.
\label{ui}
\end{equation}
Hence, from Eq. (\ref{udecomp2})
\begin{equation}
U(t) = \sum_{i=1}^N P_i(t) (u_i + \tilde{s}_i).
\label{fi1}
\end{equation}
Since $U(t)$ is the average internal energy, we must have
\begin{equation}
U(t) = \sum_{i=1}^N P_i(t) \tilde{u}_i,
\label{fi2}
\end{equation}
with $\tilde{u}_i$ the mean internal energy associated to the basin $\omega_i$. Therefore, it follows from Eqs. (\ref{fi1}) and (\ref{fi2}) that
\[
u_i = \tilde{u}_i -  \tilde{s}_i.
\]
We see from this last expression, and the fact that $\tilde{s}_i$ is an entropic term (\ref{tildes}), that $u_i$ takes the form of a conditional free energy. Finally, we have from (\ref{pi}), (\ref{udef}), and (\ref{ui}) that
\[
u_i = -\epsilon \ln \left( \int_{\omega_i} d \vx P^s(\vx,t) \right)= -\epsilon \ln \left( \int_{\omega_i} d \vx e^{- u(\vx)/k_B T} \right).
\]
in agreement with Eq. (\ref{eq_03}).

\section{Time evolution and thermodynamic process variables}

\subsection{General case}

By differentiating Eqs. (\ref{udecomp2}), (\ref{sdecomp}), and (\ref{fdecomp}) and making use of Eq. (\ref{me}) we obtain the following expressions for $\dot{U}$, $\dot{S}$, and $\dot{F}$:
\begin{eqnarray}
\dot{U} & = & \frac{\epsilon}{2} \sum_{i,j=1}^N (P_j \gamma_{ji} - P_i \gamma_{ij}) \left( \ln \frac{P_j^s}{P_i^s} - \frac{\tilde{s}_j - \tilde{s}_i}{k_B} \right), \label{dotu} \\
\dot{S} & = & \frac{\epsilon}{2} \sum_{i,j=1}^N (P_j \gamma_{ji} - P_i \gamma_{ij}) \left( \ln \frac{P_j}{P_i} - \frac{\tilde{s}_j - \tilde{s}_i}{k_B} \right), \label{dots} \\
\dot{F} & = & \frac{\epsilon}{2} \sum_{i,j=1}^N (P_j \gamma_{ji} - P_i \gamma_{ij}) \ln \frac{P_j^s P_i}{P_i^s P_j}, \label{dotf} 
\end{eqnarray}

Before proceeding any further, notice that the formulas for $\dot{U}$ and $\dot{S}$ possess both coarse- and fine-grained terms. Nonetheless, because of the adiabatic approximation, the fine-grained  terms in $\dot{U}$ and $\dot{S}$ are equal. Hence, they cancel in $U - S$ and, in consequence, the time derivative for the free energy ($\dot{F}$) is the same no matter wether the system has a fine-grained structure or not \citep{Qian:2001fk,Ge:2010fk}. 

Following a procedure completely analogous to that in \citep{Ge:2010fk} we introduce the following definitions for the entropy production rate ($e_p$), the heat dissipation rate ($Q_d$), and the housekeeping heat ($Q_{hk}$):
\begin{eqnarray}
e_p &=& \frac{\epsilon}{2} \sum_{i,j=1}^N  (P_j \gamma_{ji} - P_i \gamma_{ij}) \ln \frac{P_j \gamma_{ji}}{P_i \gamma_{ij}}, \label{sigma}  \\
Q_d &=& \frac{\epsilon}{2} \sum_{i,j=1}^N  (P_j \gamma_{ji} - P_i \gamma_{ij}) \left(\ln \frac{\gamma_{ji}}{\gamma_{ij}} + \frac{\tilde{s}_j - \tilde{s}_i}{k_B} \right),  \label{qd} \\
Q_{hk} &=&  \frac{\epsilon}{2} \sum_{i,j=1}^N  (P_j \gamma_{ji} - P_i \gamma_{ij}) \ln \frac{P_j^s \gamma_{ji}}{P_i^s \gamma_{ij}}. \label{qhk} 
\end{eqnarray}
It is straightforward to prove from the definitions above and Eqs. (\ref{dotu})-(\ref{dotf}) that
\begin{equation}
\dot{U} = Q_{hk} - Q_d, \quad
\dot{S} = e_p - Q_d, \quad
\dot{F} = Q_{hk} - e_p. 
\label{flows} 
\end{equation}

As discussed elsewhere \citep{Oono:1998ly,Ge:2010fk}, and mentioned above, $e_p$ represents the entropy production rate of the system, $Q_d$ the heat dissipation rate, and $Q_{hk}$ the energy influx rate necessary to keep the stationary distribution away from thermodynamic equilibrium.

Given that the stationary distribution satisfies (Eq. \ref{me})
\[
\sum_{j=1}^N  (P_j^s \gamma_{ji} - P_i^s \gamma_{ij}) = 0,
\]
it is not hard to prove, see \citep{Santillan:2011vn} for a detailed demonstration, that
\[
\sum_{i,j=1}^N  (P_j \gamma_{ji} - P_i \gamma_{ij}) \ln \frac{P_j^s }{P_i^s} = 0 
\quad \text{and} \quad \sum_{i,j=1}^N  (P_j \gamma_{ji} - P_i \gamma_{ij}) (\tilde{s}_j - \tilde{s}_i) = 0.
\]
These results further mean that, in the steady state,
\begin{equation}
e_p = Q_d = Q_{hk} =  \frac{\epsilon}{2} \sum_{i,j=1}^N  (P_j^s \gamma_{ji} - P_i^s \gamma_{ij}) \ln \frac{\gamma_{ji}}{\gamma_{ij}} > 0.
\label{fluxessd}
\end{equation}
That is, all fluxes are larger than zero, but they balance in such a way that $\dot{U}, \dot{S}, \dot{F} = 0$ in the steady state.

We point out that only $Q_d$ possesses a fine grained term. However, the corresponding discussion is delayed to the next subsection, in connection with the imposed adiabatic approximation and detailed balance.

\subsection{Detailed balance}

In the particular case where the stationary distribution $P^s(\vx)$ complies with detailed balance, the probability flux is null ($\vJ^s = 0$) for every $\vx$ \citep{Qian:2002fk,Kampen:2007kx,Risken:1996uq}. This, together with Eq. (\ref{sumcurr}), further implies that $J_{ij}^s=J_{ji}^s$ for all $i\neq j$. And so (Eq. \ref{linearflux}) that
\begin{equation}
\gamma_{ij} P_i^s = \gamma_{ji} P_j^s.
\label{detbal}
\end{equation}
Substitution of this last equation into Eqs. (\ref{sigma})-(\ref{qhk}) gives
\begin{eqnarray}
e_p &=& \frac{\epsilon}{2} \sum_{i,j=1}^N  (P_j \gamma_{ji} - P_i \gamma_{ij}) \ln \frac{P_j P_i^s}{P_i P_j^s}, \label{sigmadb}  \\
Q_d &=& \frac{\epsilon}{2} \sum_{i,j=1}^N  (P_j \gamma_{ji} - P_i \gamma_{ij}) \left(\ln \frac{P_i^s}{P_j^s} + \frac{\tilde{s}_j - \tilde{s}_i}{k_B} \right),  \label{qddb} \\
Q_{hk} &=&  0. \label{qhkdb} 
\end{eqnarray}
It then follows from Eq. (\ref{flows}) that
\begin{equation}
\dot{U} = - Q_d, \quad
\dot{S} = e_p - Q_d, \quad
\dot{F} = - e_p. 
\label{flowsdb} 
\end{equation}

By substituting $P_i = P_i^s$ into Eqs. (\ref{sigmadb})-(\ref{qhkdb}) and taking into account Eq. (\ref{detbal}), we have that 
\[
e_p = Q_d = Q_{hk} = 0.
\]
That is, when detailed balance is satisfied (or equivalently, when the system is in thermodynamic equilibrium), all state variables remain constant in time because all fluxes are null.

Consider again the adiabatic approximation. We can see from Eq. (\ref{qsdist}) that it is equivalent to assuming that the probability distribution immediately evolves, within each $\omega_i$, to a local quasi-stationary distribution compatible with thermodynamic equilibrium. This last fact explains why neither $e_p$ nor $Q_{kh}$ | Eqs. (\ref{sigma}) and (\ref{qhk}) | possess fine-grained terms.

\subsection{Emergent coordinate and mean-field approximation}

\begin{figure}[htb]
\includegraphics[width=2in]{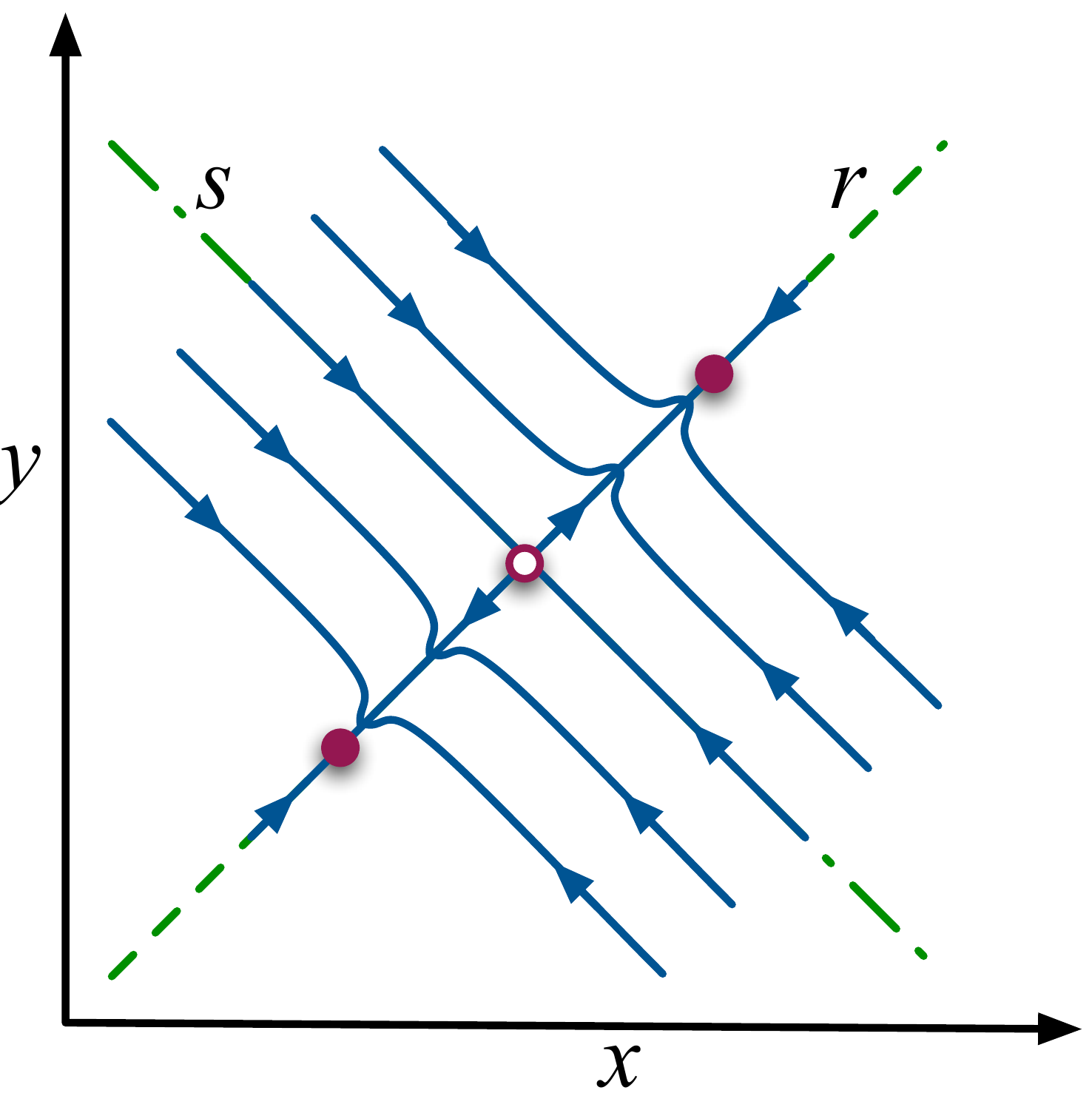}
\caption{Schematic representation of how a natural coordinate system $(r,s)$ can emerge
from the slow and fast manifolds of a given dynamical system originally described in the phase space $(x,y)$.}
\label{fig:1}
\end{figure}

The  coordinate system of the phase space of a given stochastic dynamics, $(x,y)$, usually is not the most natural one in terms of the multiscale dynamics. Fig. \ref{fig:1} illustrates how a dynamically natural coordinate system $(r,s)$ can emerge from slow and fast manifolds.  The slow manifold is widely known in chemical reaction dynamics as the ``reaction coordinate''.  The potential of mean force along the slow manifold, $A(r)$, is widely called the ``energy landscape''.
 
In terms of the emergent dynamic coordinates $(r,s)$, the partition function is
\begin{equation}
	Z(\epsilon) = \int dx \int dy\ e^{-V(x,y)/\varepsilon}
	  = \int dr\ e^{-A_r(r)/\epsilon},
\label{eq_42}
\end{equation}
in which 
\begin{equation}
		A_r(r) = -\epsilon\ln \int ds\
       \Big\|\frac{D(x,y)}{D(r,s)}\Big\|\
        e^{-\widetilde{V}(r,s)/\epsilon},
\end{equation}
with $\widetilde{V}(r,s)=V\left(x(r,s),y(r,s)\right)$. Furthermore $(r,s)$ is a coordinate transformation of $(x,y)$: $x=x(r,s),y=y(r,s)$, with a non-singular Jacobian$ \big\|{D(x,y)} / {D(r,s)}\big\|\ne 0$.

How to discover the dynamically natural slow coordinate?  One of the most widely used approaches is the {\em mean field method}. To illustrate this approach, consider the conditional free energy
\begin{equation}
	A_x(x) = -\epsilon\ln\int dy\ e^{-V(x,y)/\epsilon},
\end{equation}
and also the conditional mean value for $y$, $\langle y\rangle_x = E\left[y|x\right]$:
\begin{equation}
	  \langle y\rangle_x =
		\frac{\int dy y\ e^{-V(x,y)/\epsilon}}
			{\int dy e^{-V(x,y)/\epsilon}} 
         = e^{A_x(y) / \epsilon}\int dy y\ e^{-A_x(y)/\epsilon} . 
\end{equation}
The curve $\langle y\rangle_x$ can be considered as an emergent reaction coordinate.  Then, using $x$ as a parameter, $A_x(x)$ and $\langle y\rangle_x$ give an ``energy function'' along the reaction coodinate.  One can in fact choose a new coordinate $r$ along the curve $y=\langle y\rangle_x$. 

It is easy to verify that (see Eq. \ref{eq_42}):
\begin{equation}
	\int\ dx \; e^{-A_x(x)/\epsilon}  = \int dr \;
			e^{-A_r(r)/\epsilon} = Z(\epsilon).
\label{eq_46}
\end{equation}
All the equations so far are exact. However, in studies of real chemical and biophysical problems, one often chooses not to compute the last integral in (\ref{eq_46}).  Rather, one finds the local or global minima of $A_r(r)$.  The reasons for this practice are twofold:
\begin{itemize}
\item First, it is often analytically impossible
to carry out the integration.  In this case, finding
the global minimum ($r^*$) is a reasonable approximation,
especailly for small $\epsilon$ \cite{bender_book}:
\begin{equation}
	-\epsilon\ln \int dr e^{-A_r(r)/\epsilon} =
		A_r\left(r^*\right) -\frac{\epsilon}{2} \ln
			\left(\frac{2\pi \epsilon}{A_r''(r^*)}\right)
			+ \cdots.
\end{equation}
Since the approximation neglects the fluctuations in $r$ around $r^*$, the method is widely called {\em mean field approximation.}  In applied mathematics, this is known as Laplace's method for integrals \cite{bender_book}.   

\item Second, $A_r(r)$ might have multiple minima, say two. In that case, carrying out the integration is not as insightful as to identifying the bistability of the system, and the associated transitions.  They can be visualized by the potential of mean force
$A_r(r)$. In that case, a slow, emergent stochastic dynamics on $A_r(r)$ arises.  Both Flory-Huggins theory of polymer solutions \cite{fht_book} and the Bragg-Williams approximation for nonequilibrium steady-stat \cite{bwt_book} are successful
examples.
\end{itemize}

\section{Concluding remarks}

In this work we have studied the thermodynamic consistency, or invariance, across scales of a continuous-state continuous-time system, undergoing a Markovian stochastic process. In particular, we tackled the question of how the system thermodynamic variables, as well as the relations among them, transform when the system is described, in a coarse-grained fashion, by means of discrete variables. In that respect, we proved that, in the Helmholtz free-energy perspective, the \emph{thermodynamics} derived from the continuous underlying detailed dynamics is \emph{exact}. I.e. it is the same as if one only takes a middle-road and starts with a discrete description, with the transition rates $\gamma_{ij}$ either directly measured,  or estimated by parameter fitting of experimental data. Below we further discuss some interesting consequences from these results.

\subsection{Energy and thermodynamics across scales}

Consider a stochastic dynamical system with two levels of descriptions: an upper coarse-grained level and and a lower refined level with well-separated dynamic time scales.  Then, following the analysis in the present paper, one has $F_1 = U_1-S_1$ and $F_2=U_2-S_2$, where subscripts ``1'' and ``2'' denote upper and lower levels. Furthermore, $U_1<U_2$.  Their difference is considered to be  ``heat'' dissipated from the upper level to the lower level.

The relationship between the classical Newtonian mechanics with $S_1\approx 0$ and the molecular description of matter is an example. The energy difference $U_2-U_1$ is entropic; it can not be fully used to ``do work'' at the upper level.  The dynamics on the fast time scale at the lower level is considered to be ``fluctuations'' for the upper level. In a spontaneous transient at the upper level, ${d(U_2-U_1)}/{dt}$ is the rate of the amount of energy being passed to the lower level. Energy conservation can only be understood from the description of the lowest level. Conversely, entropy is the concept required to characterize the changing $U$ across scales.

The thermodynamics across scales in stochastic dynamics, in particular the energy dissipation from a upper scale into a lower scale, has been a central unresolved issue in the theory of turbulence \cite{frisch_book}.  Whether the newly  developed thermodynamic framework of stochastic dynamics can shed some light on the problem remains to be seen.

\subsection{Coarse-graining as conditional probability}
	
In a recent study \cite{Santillan:2011vn} we have shown that the conditional free energy, which corresponds to the potential of mean force in continuous stochastic systems, plays an essential role in the invariance of mathematical irreversible thermodynamics of multiscale stochastic systems.  Furthermore, in \cite{moy_qian_12}, we have proved that Legendre transforms between different thermodynamic potentials for different Gibbs ensembles can be derived in terms of conditional probability for a pair of random variables. As a matter of fact, one can consider coarse-graining as a special form of conditional probability. Treating $\{(\vx,i)|\vx\in\mathbb{R}^M,1\le i\le N\}$ as a pair of random variables, the coase-graining means
\begin{equation}
       f(\vx|\ell) = \frac{\Pr\{ x\le\vx\le x+d\vx|i=\ell\}}
                {d\vx} =\left\{
          \begin{array}{ccc}
             {f_{\ell}(\vx)} / {P_{\ell}} 
             &&  \vx\in\omega_{\ell}, \\
             0 &&  \vx\notin\omega_{\ell},
          \end{array} \right.
\end{equation}
in which $f_{\ell}(\vx)$ is the joint probability,  $f(\vx|\ell)$ is the conditional probability, and 
\begin{equation}
    P_{\ell} = \int_{\Omega} d \vx f_{\ell}(\vx) = \int_{\omega_{\ell}}d \vx
               f_{\ell}(\vx).
\end{equation}
Then, the standard chain rule for free energy (i.e. relative entropy) \cite{cover_book},
\begin{equation}
   \sum_{\ell=1}^N\int  _{\Omega} d \vx f_{\ell}(\vx)\ln
            \frac{f_{\ell}(\vx)}{f_{\ell}^{s}(\vx)}
   = \sum_{\ell=1}^N P_{\ell}\ln\frac{P_{\ell}}{P^s_{\ell}}
   +  \sum_{\ell=1}^N P_{\ell}
           \left(\int_{\Omega} d \vx f(\vx|\ell)\ln
            \frac{f(\vx|\ell)}{f^{s}(\vx|\ell)}
       \right),
\label{total_f}
\end{equation}
takes an interesting, equivalent form: 
\begin{equation}
  \sum_{\ell=1}^N P_{\ell}\ln\frac{P_{\ell}}{P^s_{\ell}}
   +  \sum_{\ell=1}^N P_{\ell}
           \left(\int_{\omega_{\ell}} d \vx f(\vx|\ell)\ln
            \frac{f(\vx|\ell)}{f^{s}(\vx|\ell)}
       \right),
\label{eq_51}
\end{equation}
in which 
\[ 
   \int_{\omega_{\ell}} d \vx f(\vx|\ell)\ln
            \frac{f(\vx|\ell)}{f^{s}(\vx|\ell)}
\]
is the ``conditional free energy'' of the sub-system $\ell$. For subsystems with rapid steady state, this term is 
zero. Thus, a system's total free energy (\ref{total_f}) is the free energy of the coase-grained system.

\acknowledgments

The authors are in debt with Prof. Eduardo S. Zeron for the proof in Appendix \ref{boundary}.

\bibliography{DiscreteApprox}

\appendix

\section{The entropy depends on how finely the system is described}
\label{appendix}

Consider the definitions for $S(\epsilon)$ and $\widetilde{S}(\epsilon)$ given in Eq. (\ref{EntRef}), and rewrite them as
\[
S(\epsilon) = - \epsilon \sum_i p_i \ln p_i, \quad \text{and} \quad 
\widetilde{S}(\epsilon) = -\epsilon \int_{\Omega} d \vx \rho(\vx) \ln \rho(\vx),
\]
with
\[
p_i = \frac{e^{-A_i / \epsilon}}{Z(\epsilon)}, \quad \text{and} \quad \rho(\vx) = \frac{e^{-V(\vx) / \epsilon}}{\widetilde{Z}(\epsilon)}.
\]
It follows from Eqs. (\ref{eq_01})-(\ref{eq_03}) and the former definitions that
\[
p_i = \int_{\omega_i} d \vx \rho(\vx),
\]
in which $\bigcup_{i=1}\omega_i=\Omega$.
We can now use this last result to rewrite $\widetilde{S}(\epsilon)$ as
\[
\widetilde{S}(\epsilon) = S(\epsilon) -\epsilon  \sum_i p_i \int_{\omega_i} d \vx
\left(\frac{\rho(\vx)}{p_i}\right) \ln \left(\frac{\rho(\vx)}{p_i}\right) .
\] 
However, since $\int_{\omega_i} d \vx \rho(\vx)/p_i  = 1$,
\[
\widetilde{S}_i(\epsilon) \triangleq  -\epsilon \int_{\omega_i} d \vx
\left(\frac{\rho(\vx)}{p_i}\right)\ln\left(\frac{\rho(\vx)}{p_i}\right) \, 
 \geq 0
\]
is a conditional entropy associated to the probability distribution within 
$\omega_i$. Hence,
\[
\widetilde{S}(\epsilon) - S(\epsilon) = \sum_i p_i \widetilde{S}_i(\epsilon) \geq 0.
\]

\section{Analysis of the boundaries of the regions $\omega_i$ covering $\Omega$}
\label{boundary}

We start with some definitions. Let $X$ be an arbitrary set in $\mathbb{R}^n$. The closure of $X$, $Cl(X)$, is defined as the intersection of all closed sets $C$ such that $X \subset C$. On the other hand, the interior of $X$, $In(X)$, is defined as the union of all open sets $A$ such that $A \subset X$. Finally, the boundary of $X$, $Bd(x)$, is defined as $Bd(X) = Cl(X) \setminus In(X)$.

Let $\Omega \in \mathbb{R}^n$ be an open set and $\{\omega_i\}$ a cover of $\Omega$ such that $\omega_i \subset \Omega$ for all $i$, and $\Omega = \bigcup_{i=1}^N \omega_i$. We make the following assertions regarding $\Omega$ and $\{\omega_i\}$:

\begin{enumerate}
\item Observe that, since the union of closed sets is closed, $\Omega \subset \bigcup_{i=1}^N Cl(\omega_i)$, with $\bigcup_{i=1}^N Cl(\omega_i)$ a closed set. Furthermore, $Cl(\Omega) \subset \bigcup_{i=1}^N Cl(\omega_i)$

\item Note also that $\omega_i \supset \left[ \Omega \setminus \bigcup_{j \neq i} \omega_j \right]$. 

\item Moreover, given that $\Omega$ is open, $\omega_i \supset D$, with $D = \Omega \setminus \bigcup_{j \neq i} Cl(\omega_j)$ an open set. And so, $D \subset In(\omega_i)$.

\item From the definitions above, the boundary of $\omega_i$, $\Xi_i = Bd(\omega_i)$, is a closed set contained in $Cl(\Omega)$. Hence, from Assertion 3, $In(\omega_i) = D \bigcup In(\omega_i)$ and, from the definition of a set boundary, 
\begin{eqnarray}
\Xi_i & = & Cl(\omega_i) \setminus In(\omega_i), \nonumber \\
 & = & Cl(\omega_i) \setminus \left[D \bigcup In(\omega_i)\right], \nonumber \\ 
 & = & [Cl(\omega_i) \setminus In(\omega_i)] \bigcap [Cl(\omega_i) \setminus D], \nonumber \\
 & = & \Xi_i \bigcap [Cl(\omega_i) \setminus D], \nonumber \\ 
 & = & \left[\Xi_i \bigcap Cl(\omega_i)\right] \setminus D, \nonumber \\
 & = & \Xi_i \setminus D. \nonumber \\
\end{eqnarray}

\item Let us define now $H = \bigcup_{j \neq i} Cl(\omega_j)$, so $D = \Omega \setminus H$. From this:
\begin{eqnarray}
\Xi_i & = & \Xi_i \setminus D, \nonumber \\
 & = & \Xi_i \setminus [\Omega \setminus H], \nonumber \\ 
 & = & [\Xi_i \setminus \Omega] \bigcup [\Xi_i \bigcap H]. \nonumber \\
\end{eqnarray}

\item We have from the definition of $H$ that 
\[
\Xi_i = [\Xi_i \setminus \Omega] \bigcup\left[\bigcup_{i\neq j} [\Xi_i \bigcap Cl(\omega_j)]\right].
\]
Recall that $\Omega$ is open, so $\Xi_i \setminus \Omega$ is the part of the closed set $\Xi_i$ lying outside $\Omega$. On the other hand, $\Xi_i \bigcap Cl(\omega_j)$ is the part of $\Xi_i$ lying within the closure of $\omega_j$. 

\item Notice that 
\begin{eqnarray}
\Xi_i \bigcap \Xi_j & = & \Xi_i \bigcap [Cl(\omega_j) \setminus In(\omega_j)], \nonumber \\
 & = & \left[\Xi_i \bigcap Cl(\omega_j)\right] \setminus In(\omega_j), \nonumber \\
 & = & Cl(\omega_j) \bigcap [\Xi_i \setminus In(\omega_j)]. \nonumber
\end{eqnarray}

\item Furthermore, 
\begin{eqnarray}
\Xi_i \setminus In(\omega_j) & = & [Cl(\omega_i) \setminus In(\omega_i)] \setminus In(\omega_j), \nonumber \\
 & = & [Cl(\omega_i) \setminus In(\omega_j)] \setminus In(\omega_i). \nonumber
\end{eqnarray}

\item If we assume now that $Cl(\omega_i)$ and $In(\omega_j)$ are disjoint, then $Cl(\omega_i) \setminus In(\omega_j) = Cl(\omega_i)$. Then, from Assertion 8,
\[
\Xi_i \setminus In(\omega_j) = Cl(\omega_i) \setminus In(\omega_i) = \Xi_i.
\]
Moreover, from Assertion 7, $\Xi_i \setminus In(\omega_j) = Cl(\omega_i) \bigcap \Xi_i$
\end{enumerate}

In conclusion, if we assume that $\Omega \in \mathbb{R}^n$ is an open set such that $\Omega = \bigcup_i \omega_i$, and we assume as well that $Cl(\omega_i)$ and $In(\omega_j)$ are disjoint for all $i \neq j$ then, from Assertion 6,
\[
\Xi_i = [\Xi_i \setminus \Omega] \bigcup \left[ \bigcup_{i \neq j} [\Xi_i \bigcap \Xi_j ] \right].
\]

Denote $\Xi_{i0} = \Xi_i \setminus \Omega$ and $\Xi_{ij} = \Xi_i \bigcap \Xi_j $, so
\[
\Xi_i = \bigcup_{j=0}^n \Xi_{ij}.
\]
\qed

\end{document}